# Optical computing with supercontinuum generation in photonic crystal fibers

AZKA MAULA ISKANDAR MUDA, UĞUR TEĞIN*

*Dept. of Electrical and Electronics Engineering, Koç University*
*\*utegin@ku.edu.tr*

**We introduce a novel photonic neural network using photonic crystal fibers, leveraging femtosecond pulse supercontinuum generation for optical computing. Investigating its efficacy across machine learning tasks, we uncover the crucial impact of nonlinear pulse propagation dynamics on network performance. Our findings show that octave-spanning supercontinuum generation results in loss of dataset variety due to many-to-one mapping, and optimal performance requires balancing optical nonlinearity. This study offers guidance for designing energy-efficient and high-performance photonic neural network architectures by explaining the interplay between nonlinear dynamics and optical computing.**

Machine learning, and particularly deep neural networks, have achieved remarkable successes in a variety of applications, from image recognition and natural language processing to object detection and autonomous driving [1]. However, the increasing complexity of these models has led to significant challenges in terms of computational expense and environmental impact. As machine learning models grow larger and more intricate, they require vast amounts of data and computational resources for training [2]. Traditional Turing-von Neumann architecture-based digital computing tools are struggling to keep up with these demands. Moreover, the energy consumption of these large machine learning models is becoming a major concern. With the introduction of generative models like GPT-3, which can generate human-like text on demand, the energy requirements have started to raise eyebrows [3]. Current estimates suggest that generating just one image with a generative AI model uses as much energy as charging an average cellphone [4]. Therefore, there is a pressing need for more energy-efficient and environmentally friendly machine learning architectures.

Optical computing, a subfield of analog computing, has gained considerable attention in recent years for its potential applications in machine learning studies [5]. It offers several advantageous operations, such as Fourier transform, convolution, and matrix multiplication, which can be achieved intrinsically using passive components. The inherent robustness of deep neural networks against noise makes them an ideal candidate for optical computing research. The concept of performing machine learning calculations at the speed of light while conserving energy is an intriguing subject and was first proposed in 1985 with the implementation of an optical Hopfield network [6]. Since then, various techniques have been explored to achieve this goal, including volume holograms [7,8], nanophotonic circuits [9,10], and light modulators [11]. The development of neuromorphic computing tools like reservoir learning [12-14] helped overcome this obstacle by providing a simple and robust approach for training optical neural networks. It led to the introduction of diffraction and electrooptic components-based optical neural network architectures [15,16]. However, a major challenge in the field of optical computing for machine learning applications remained the implementation of complex nonlinear transfer functions to introduce dimensionality expansion to datasets while processing information optically.

Nonlinear propagation of light in matter has emerged as a promising tool for performing nonlinear optical computing for machine learning applications. This concept relies on utilizing the nonlinear properties of light as it interacts with and propagates through various types of materials to perform complex computations. One of the earliest theoretical models proposing this approach was presented by Marcucci et al., who suggested using the normalized nonlinear Schrödinger equation and soliton dynamics as a reservoir computing platform for machine learning tasks [17]. Later, spatiotemporal nonlinear propagation of light in multimode optical fibers are utilized to demonstrate scalable nonlinear optical computing to spatial information carrying laser pulses in graded-index multimode fibers for machine learning applications [18] and an programmable approach is demonstrated recently [19]. In recent years, research focused on using single mode fibers and lithium niobate based waveguides for nonlinear feature generation in machine learning applications [20,21].

In this letter, we propose and study the coherent supercontinuum generation with photonic crystal fibers (PCFs) as an optical computing platform. By employing the underlying dynamics of octave-spanning supercontinuum generation as the basis of nonlinear optical computing, we investigate the performance of our PCF-based photonic neural network. We investigate the effect of nonlinear pulse propagation of the information-carrying femtosecond pulses in PCF with detailed state-of-the-art tools. Our study demonstrates that optical nonlinearity and spectral broadening must be adjusted to achieve high-performance machine learning with our PCF-based photonic neural network. We employ statistical tools like linear discriminant analysis (LDA) to visualize how different classes of data are separated by the dimensionality expansion provided by optical nonlinear dynamics. Our results show that the generation of octave-spanning supercontinua results in many-to-one mapping and limits the performance of the photonic neural network. By systematically varying the peak powers and propagation distances across three different datasets, namely Sinc regression, iris, and liver disease classification, our study shows the need for balance between nonlinear pulse propagation dynamics of the fiber and machine learning performance.

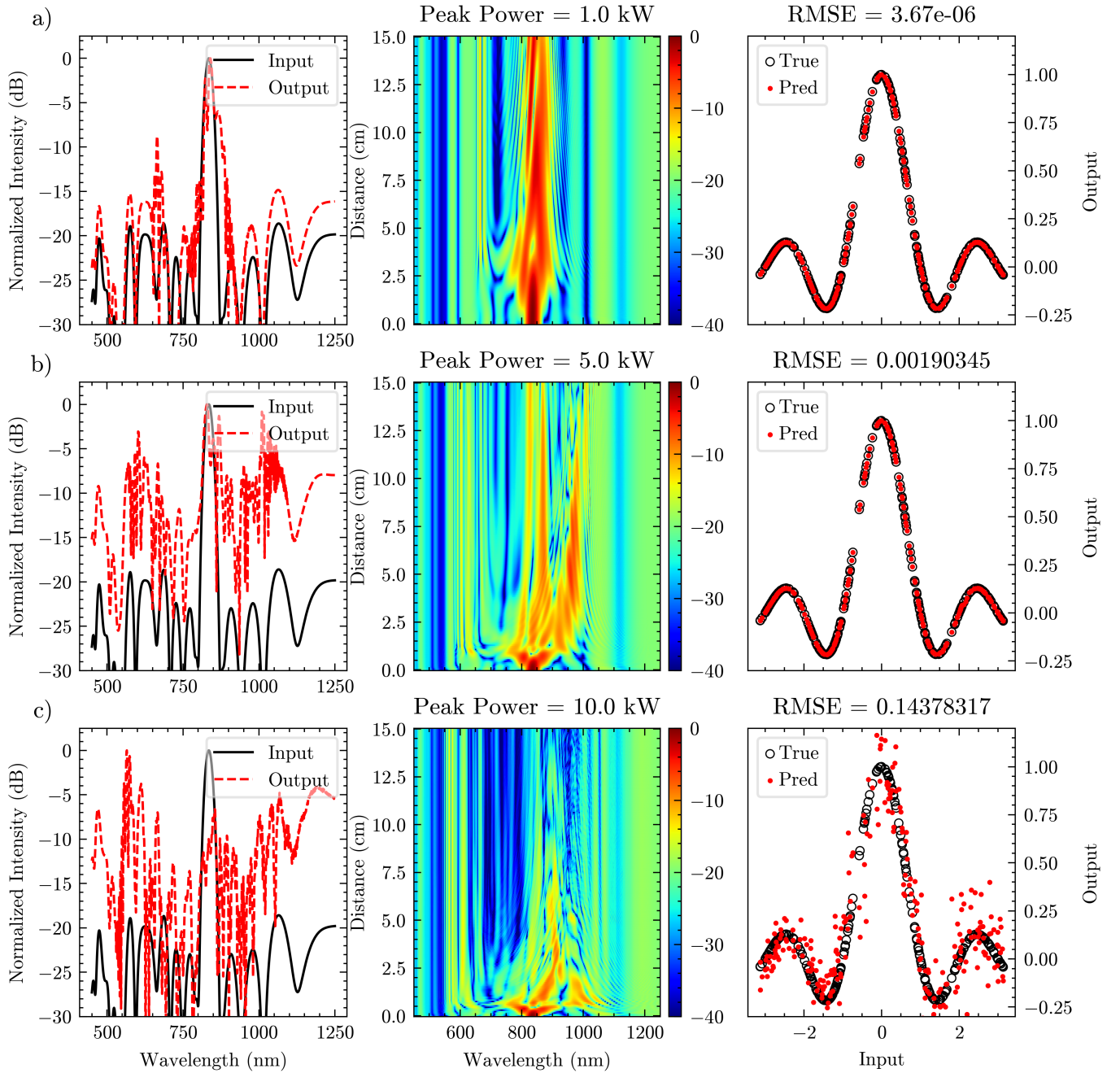

Fig. 1. Sinc dataset results with PCF-based photonic neural network. For pulses with 1 kW (a), 5 kW (b) and 10 kW (c) peak powers typical input and output spectra, spectral evolution and linear regression results with nonlinear optical computing.

Supercontinuum generation in PCFs has been studied extensively in recent decades, and numerical tools for studying the underlying complex dynamics to generate supercontinua have matured [22-25]. Our analysis is based on state-of-the-art numerical simulations of the nonlinear Schrödinger equation with the 4th order Runge-Kutta in the interaction picture (RK4IP) algorithm [26]. We study

$$\frac{\partial A}{\partial z} = \sum_{k \geq 2}^{10} \frac{i^{k+1}}{k!} \beta_k \frac{\partial^k A}{\partial T^k} + i\gamma(|A|^2 A + \frac{i}{\omega_0}\frac{\partial}{\partial T}(|A|^2 A) - T_R A \frac{\partial |A|^2}{\partial T})$$

where $A(z, t)$ is the time-domain envelope of the electric field, $z$ is the propagation distance, $T$ is the time, $\beta_k$ are the dispersion coefficients, $\gamma$ is the nonlinear coefficient, $\omega_0$ is the central frequency, and $T_R$ is the Raman response time of the fiber. A photonic crystal fiber, characterized by parameters given by Dudley et al. [23], is utilized in this study as the test fiber. The central wavelength of the input pulse is set at 835 nm, with the pulse enveloping a Gaussian shape characterized by a full width at half maximum (FWHM) of 50 femtoseconds (fs). The central wavelength is defined according to the zero dispersion wavelength of the PCF to realize coherent supercontinuum generation. In our studies, a temporal window of 25 fs and a discretization comprising $2^{14}$ linearly spaced points are selected. With this powerful numerical tool, we investigate spectral broadening of the femtosecond high-power pulses for the fiber lengths of 10 cm, 12 cm, and 15 cm with different peak powers. We set the step size in our simulations as 30 $\mu$m.

Our study consists of an investigation of testing the PCF-based photonic neural network's performance with Sinc regression, iris classification [27], and liver disease classification [28] datasets. Each dataset is encoded into the input pulse through phase modulation. A data size expansion is achieved by multiplying the input with a randomly generated phase mask to ensure the use of the same amount of data points on phase modulation. After the nonlinear propagation of information-carrying pulses in PCFs, an average binning is applied with a 1 nm resolution before decoding the output spectrum. The output spectra contain 800 data points between 450-1250 nm wavelength range. The decoding algorithm is then applied to the output spectrum for the completion of the photonic neural network platform. The decoding algorithm is trained and tested on separate training and testing sets. The performance of the decoding algorithm is assessed through the root mean square error (RMSE) for the regression task and the accuracy score for the classification tasks.

The Sinc regression dataset is constructed from 500 linearly spaced data points ranging from $-\pi$ to $\pi$. The dataset is later equally divided into training and testing sets. Linear regression served as the decoding mechanism for this dataset. The iris dataset, derived from a publicly available source, is divided into equal parts for the training and test sets. The liver disease dataset underwent a synthetic minority oversampling technique (SMOTE) to ensure balanced representation within the dataset. This dataset is split into a 70% training set and a 30% testing set. For the classification tasks related to iris and liver disease, a Linear Support Vector Classifier (SVC) is adopted as the decoding algorithm. The optimization of the SVC's regularization parameter is achieved through a grid search cross-validation strategy, employing a five-fold (k = 5) approach. The performance of the classification models is assessed through the confusion matrices and accuracy scores. Furthermore, Linear Discriminant Analysis (LDA) is utilized as a means of visualizing the classification outcomes.

We start our tests with a relatively simple Sinc dataset. If a linear regression is applied to the Sinc dataset, a linear line in the middle emerges with an RMSE of 0.35. Our PCF-based photonic neural network consists of an encoding layer, coherent supercontinuum generation dynamics, and decoding layers (linear regression) with an absolute squared function due to intensity measurements. Our optical computing platform works exceptionally well for regression tasks for the Sinc dataset (see Fig. 1). It indicates that nonlinear pulse propagation in PCFs nonlinearly maps the input data to a Z-space where linear regression can be performed on the data well. At peak power of 1 kW, the photonic neural network is able to perform regression on the dataset with a remarkably low RMSE of $3.67 \times 10^{-6}$. As the peak power increases, a notable increase in RMSE is also observed. By 5 kW peak power, the RMSE increases by 3 orders of magnitude into $1.9 \times 10^{-3}$, compared to the 1 kW case. The error is even more significant with 10 kW peak power, reaching RMSE larger than 0.14. The accumulated nonlinearity of the system is also evident in the spectral evolution of the system, as shown in Fig. 1. For a fixed fiber length, the spectral evolution of the system is strongly correlated with the peak power, with the spectral broadening and the generation of supercontinuum being more pronounced as the peak power increases. Our observation underlines a counter-intuitive relation between the strength of the nonlinear dynamics and the performance of the PCF-based optical computing system in the context of neuromorphic computing.

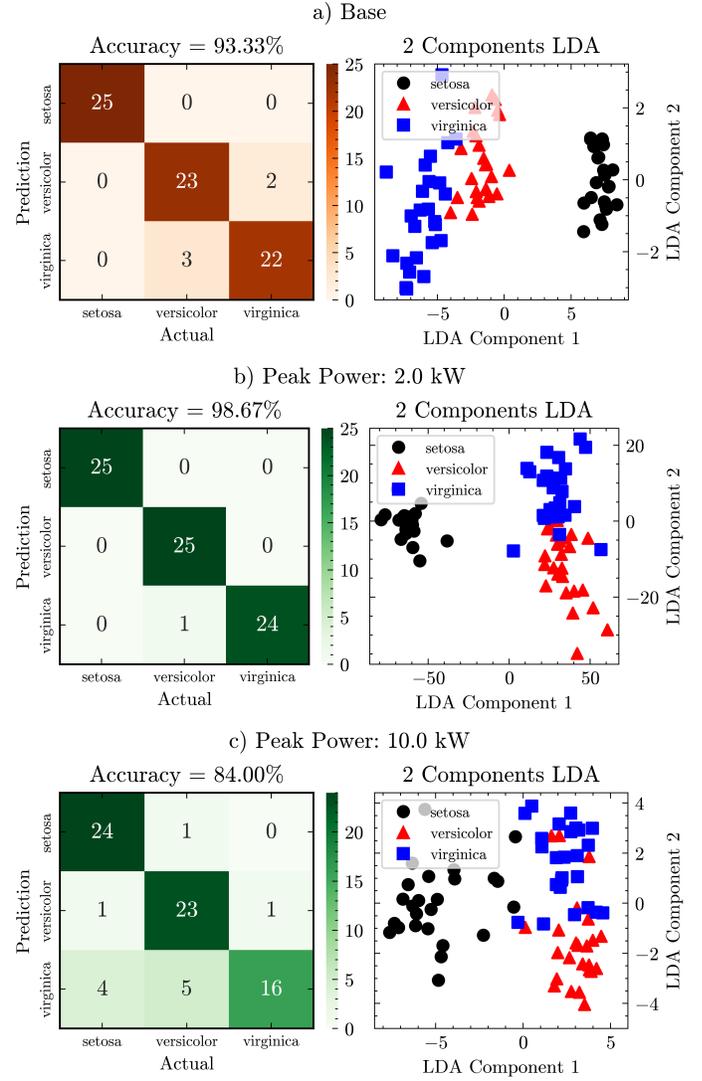

Fig. 2. Iris dataset results with PCF-based photonic neural network. Confusion matrices and LDA results without optical computing (a) and with 2 kW (b) and 5kW (b) peak powers.

Later, we study our photonic neural network performance for a classification task with the iris dataset. Instead of linear regression after the nonlinear propagation of light, we use a linear SVC to classify the samples in the dataset. To set a baseline, this linear SVC is applied to the dataset without optical computing, and a classification accuracy of 93.33% is observed. With optical computing, the classification accuracy increases by 5.33% and reaches 98.67% for a peak power of 2.0 kW and 10 cm fiber length. The confusion matrix and LDA of the iris dataset are shown in Fig. 2. As demonstrated by the confusion matrices, the system is able to classify the iris dataset with very minimal error. The LDA also shows that the dataset is linearly separable, with few outliers. However, when the peak power is increased to 10.0 kW, similar to the Sinc dataset, the classification performance of the photonic neural network suffers.

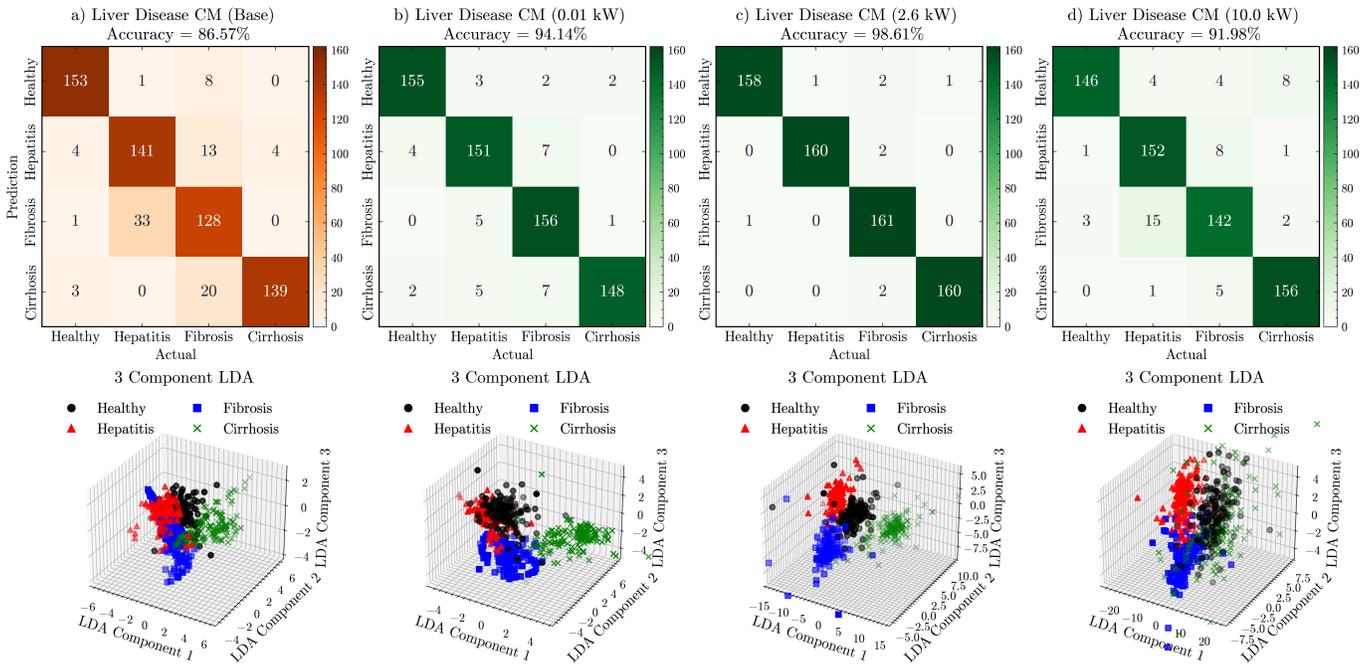

Fig. 3. Liver disease dataset confusion matrix and LDA without optical computing (a) and with 0.01 kW (b), 2.7 kW(c), and 10 kW (d) peak powers.

Encouraged by the performance we obtained with the relatively simple tasks described so far, we test our optical computing platform with the liver disease dataset. The confusion matrices and LDAs of the liver disease dataset are shown in Fig. 3. For the base case, the classification accuracy of 86.57% is achieved with only Linear SVC (see Fig. 3 a). For 10 W peak power, where the propagation can be considered linear, the encoding and decoding schemes, as well as the intensity measurement, raise the classification accuracy to 94.29%. By increasing the peak power to 2.6 kW, the nonlinear dynamics in PCF help the machine learning tasks. The accuracy of the photonic neural network reaches 98.61%, which signifies a 12.04% increase compared to the base case. The LDA shows that the dataset is clustered, with less overlap between the classes, compared to the base case (see Fig. 3 c). When the peak power is increased to 10 kW, where single octave supercontinuum is generated, the accuracy of the model deteriorates to 91.98%, which is smaller compared to the linear propagation. The LDA for 10 kW peak power indicates that classes of the dataset start to mix again, and this causes the decoding algorithm to struggle to distinguish the classes.

To assess the effect of nonlinear dynamics in PCF for femtosecond pulses on the performance of the photonic neural network, we investigate the impact of the peak power to training and test accuracy of the liver disease dataset for the fiber lengths of 10 cm, 12 cm, and 15 cm. The evolution of the train and test accuracy of the liver disease dataset is shown in Fig. 4. It can be observed that the train and test accuracy of the model increases up to 2.7 kW. Beyond this point, the train accuracy has an increasing trend, but the test accuracy of the model deteriorates. This suggests that the model is overfitting the training data when the accumulated nonlinearity of the system is high. However, when a supercontinuum is formed, both training and test accuracy of the model decreases. This is due to the many-to-one mapping conditions in the supercontinuum generation, which limits the performance of the photonic neural network.

In our study, we observe that the strength of nonlinearity has to be adjusted in supercontinuum generation to reach better results in machine learning applications. For a fixed fiber length, it means that the peak power of the femtosecond pulses, which carry information, has to be finetuned. Our analyses show that when the peak power is above a certain value, capable of generating an octave-spanning supercontinuum spectrum for the sample, the performance of the photonic neural network suffers drastically. We observe that when the optical nonlinearity is high, the information-carrying input pulses evolve to octave-spanning supercontinua, and it creates many-to-one mapping conditions. Regardless of the input data, supercontinua at the output features a similar wavelength distribution due to the saturation of the nonlinear dynamics. Such a mapping destroys the variety in the dataset, and the performance of the photonic neural network decreases. When the optical nonlinearity is adjusted below the threshold of supercontinuum generation, the nonlinear dynamics help to differentiate the small differences between the samples and increase dimensionality. A similar relation has been recently demonstrated in analog computing with Lorenz attractor [29].

In conclusion, our investigation highlights the promising potential of optical computing with PCFs and coherent supercontinuum generation as a platform for machine learning tasks. We have demonstrated the effectiveness of our PCF-based photonic neural network through comprehensive studies involving regression and classification datasets.

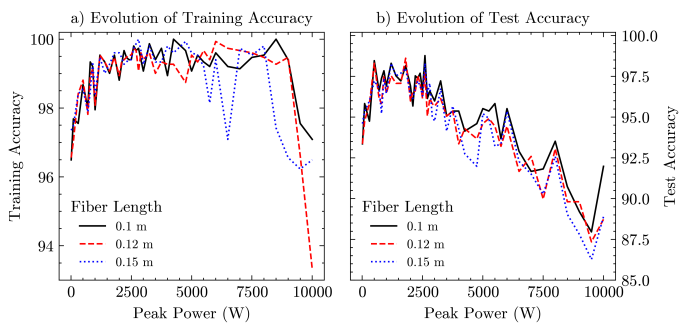

Fig. 4. Evolution of the test and cross-validation accuracy for the liver disease dataset for increasing peak power with 10 cm, 12 cm and 15 cm PCFs.

Our study reveals that the nonlinear pulse propagation dynamics in PCFs play a crucial role in shaping the performance of the photonic neural network. Specifically, we observe that optimal performance is achieved when balancing the strength of optical nonlinearity to avoid the detrimental effects of octave-spanning supercontinuum generation. Above a certain threshold of optical nonlinearity, the network experiences decreased performance due to the loss of dataset variety caused by many-to-one mapping conditions. Our results underscore the necessity for fine-tuning optical nonlinearity to maintain an effective balance between dimensionality expansion and model performance. Furthermore, our study provides insights into the intricate interplay between optical dynamics and machine learning tasks, offering valuable guidance for designing and optimizing future photonic neural network architectures.

**Funding.** We received no specific funding for this work.

## References

1. E. Alpaydin, Machine learning (MIT Press, 2021).
2. G. Team, R. Anil, S. Borgeaud, Y. Wu, J.-B. Alayrac, J. Yu, R. Soricut, J. Schalkwyk, A. M. Dai, A. Hauth, et al., Gemini: a family of highly capable multimodal models, arXiv preprint arXiv:2312.11805 (2023).
3. G. Tamburrini, The ai carbon footprint and responsibilities of ai scientists, Philosophies 7, 4 (2022).
4. Melissa Heikkila, Making an image with generative ai uses as much energy as charging your phone (2023), accessed Feb 10, 2024. https://www.technologyreview.com/2023/12/01/1084189/ making-an-image-with-generative-ai-uses-as-much/-energy-as-charging-your-phone/.
5. G. Wetzstein, A. Ozcan, S. Gigan, S. Fan, D. Englund, M. Soljačić, C. Denz, D. A. Miller, and D. Psaltis, Inference in artificial intelligence with deep optics and photonics, Nature 588, 39 (2020).
6. N. H. Farhat, D. Psaltis, A. Prata, and E. Paek, Optical implementation of the hopfield model, Applied optics 24, 1469 (1985).
7. Y. S. Abu-Mostafa and D. Psaltis, Optical neural computers, Scientific American 256, 88 (1987).
8. D. Psaltis, D. Brady, and K. Wagner, Adaptive optical networks using photorefractive crystals, Applied Optics 27, 1752 (1988).
9. Y. Shen, N. C. Harris, S. Skirlo, M. Prabhu, T. Baehr-Jones, M. Hochberg, X. Sun, S. Zhao, H. Larochelle, D. Englund, et al., Deep learning with coherent nanophotonic circuits, Nature photonics 11, 441 (2017).
10. N. Mohammadi Estakhri, B. Edwards, and N. Engheta, Inverse-designed metastructures that solve equations, Science 363, 1333 (2019).
11. J. Spall, X. Guo, T. D. Barrett, and A. Lvovsky, Fully reconfigurable coherent optical vector–matrix multiplication, Optics Letters 45, 5752 (2020).
12. G.-B. Huang, Q.-Y. Zhu, and C.-K. Siew, Extreme learning machine: theory and applications, Neurocomputing 70, 489 (2006).
13. W. Maass, Liquid state machines: motivation, theory, and applications, Computability in context: computation and logic in the real world, 275 (2011).
14. C. Gallicchio and A. Micheli, Tree echo state networks, Neurocomputing 101, 319 (2013).
15. S. Gigan, Imaging and computing with disorder, Nature Physics 18, 980 (2022).
16. B. Çarpınlıoğlu, B. S. Daniş, and U. Teğin, Genetically programmable optical random neural networks. Optica Open. 10.1364/opticaopen.25040585.v1 (2024).
17. G. Marcucci, D. Pierangeli, and C. Conti, Theory of neuromorphic computing by waves: machine learning by rogue waves, dispersive shocks, and solitons, Physical Review Letters 125, 093901 (2020).
18. U. Teğin, M. Yıldırım, İ. Oğuz, C. Moser, and D. Psaltis, Scalable optical learning operator, Nature Computational Science 1, 542 (2021).
19. İ. Oğuz, J.-L. Hsieh, N. U. Dinç, U. Teğin, M. Yıldırım, C. Gigli, C. Moser, and D. Psaltis, Programming nonlinear propagation for efficient optical learning machines, Advanced Photonics 6, 016002 (2024).
20. M. Yıldırım, İ. Oğuz, F. Kaufmann, M. R. Escale, R. Grange, D. Psaltis, and C. Moser, Nonlinear optical feature generator for machine learning, APL Photonics 8 (2023).
21. B. Fischer, M. Chemnitz, Y. Zhu, N. Perron, P. Roztocki, B. MacLellan, L. Di Lauro, A. Aadhi, C. Rimoldi, T. H. Falk, et al., Neuromorphic computing via fission-based broadband frequency generation, Advanced Science 10, 2303835 (2023).
22. J. M. Dudley and S. Coen, Numerical simulations and coherence properties of supercontinuum generation in photonic crystal and tapered optical fibers, IEEE Journal of selected topics in quantum electronics 8, 651 (2002).
23. J. M. Dudley, G. Genty, , and S. Coen, Supercontinuum generation in photonic crystal fiber, Reviews of Modern Physics 78 (2006).
24. J. M. Dudley and J. R. Taylor, Ten years of nonlinear optics in photonic crystal fibre, Nature Photonics 3, 85 (2009).
25. J. M. Dudley and J. R. Taylor, Supercontinuum generation in optical fibers (Cambridge university press, 2010).
26. J. Hult, A fourth-order runge–kutta in the interaction picture method for simulating supercontinuum generation in optical fibers, Journal of Lightwave Technology 25 (2007).
27. R. A. Fisher, Iris, UCI Machine Learning Repository (1988), DOI: https://doi.org/10.24432/C56C76.
28. R. Lichtinghagen, F. Klawonn, and G. Hoffmann, HCV data, UCI Machine Learning Repository (2020), DOI: https://doi.org/10.24432/C5D612.
29. B. U. Kesgin and U. Teğin, Machine learning with chaotic strange attractors, arXiv preprint arXiv:2309.13361 (2023).